\documentclass{aip-cp}

\usepackage[numbers]{natbib}
\usepackage{rotating}
\usepackage{graphicx}
\usepackage{bm}

\def\gtap{\ \raise.3ex\hbox{$>$\kern-.75em\lower1ex\hbox{$\sim$}}\ }
\def\ltap{\ \raise.3ex\hbox{$<$\kern-.75em\lower1ex\hbox{$\sim$}}\ }

\begin{document}

% Title portion
\title{$Z_c(4430)$, $Z_c(4200)$,
$Z_1(4050)$, and $Z_2(4250)$\\
 as triangle singularities}

\author[aff1,aff2]{Satoshi X. Nakamura\corref{cor1}}

\affil[aff1] {University of Science and Technology of China, Hefei 230026, 
People's Republic of China}
\affil[aff2]{State Key Laboratory of Particle Detection and Electronics (IHEP-USTC), Hefei 230036, People's Republic of China}
\corresp[cor1]{Corresponding author: satoshi@ustc.edu.cn}

\maketitle

  \begin{abstract}
   $Z_c(4430)$ discovered in $\bar{B}^0\to\psi(2S)K^-\pi^+$,
   $Z_c(4200)$ found in $\bar{B}^0\to J/\psi K^-\pi^+$,
   and $Z_1(4050)$ and $Z_2(4250)$
observed in $\bar{B}^0\to\chi_{c1}K^-\pi^+$ 
   are candidates of charged charmonium-like states.
All surviving theoretical models
interpreted these candidates as four-quark states, until
we recently identified a compelling alternative.
We discuss that kinematical singularities in triangle loop diagrams induce
a resonance-like behavior that can consistently explain the
 properties (such as spin-parity, mass, width, and Argand plot) of 
 $Z_c(4430)$, $Z_c(4200)$, $Z_1(4050)$ and $Z_2(4250)$
from experiments.
In terms of the triangle singularities, we can also naturally
understand interesting experimental findings such as 
the appearance (absence) of $Z_c(4200)$($Z_c(4430)$)-like
contribution in  $\Lambda_b^0\to J/\psi p\pi^-$,
and the highly asymmetric shape of 
the spectrum bump for $Z_1(4050)$;
the other theoretical models have not successfully addressed these
points. 
Although Pakhlov et al. proposed  
another triangle diagram to generate
a $Z_c(4430)$-like bump,
we argue that this scenario is very unlikely.
  \end{abstract}

% Head 1
\section{Introduction}

A current trend of the hadron spectroscopy is to establish the existence
and the internal structure of exotic hadrons which are not accommodated
by the conventional $q\bar q$ and $qqq$ structures.
Such exotic hadrons could be tetraquark, pentaquark, hadron molecule, or
hybrid states.
Possible experimental signatures of the exotic states are:
(i) the mass does not fit a quark-model prediction;
(ii) the state matches a state for which Lattice QCD predicts a high
gluon content;
(iii) the state has a peculiar decay pattern;
and so on.
But the listed signatures may seem model-dependent criteria,
and one may wonder if there is a more unambiguous signature. 

The discoveries of charged quarkonium-like state candidates,
$Z_c$ and $Z_b$, are encouraging.
For example,
$Z_c(4430)$ was discovered in the 
$\psi(2S)\pi^+$ invariant mass distribution
of $\bar B^0\to\psi(2S) K^-\pi^+$~\cite{belle_z4430_2008,belle_z4430,lhcb_z4430},
while $Z_c(4200)$ in the $J/\psi\pi^+$ distribution of
$\bar B^0\to J/\psi K^-\pi^+$~\cite{belle_z4200}.
$Z_1(4050)$ and $Z_2(4250)$ are also reported in the analysis of 
$\bar{B}^0\to\chi_{c1}K^-\pi^+$~\cite{belle_z4050}.
If these spectrum bumps are really associated with the existence of
resonances, the quark content of these states minimally need four quarks, a
clear signature of exotics.
Among the charged charmonium-like states, 
$Z_c(4430)$ has been an outstanding exotic candidate~\cite{z4430-web},
and all the surviving theoretical interpretations of $Z_c(4430)$ 
considered it to be a genuine four-quark state (including
hadron-molecule interpretations) until the present work. 
Theoretical interpretations of $Z_c(4200)$, $Z_1(4050)$ and $Z_2(4250)$
are similar, although these exotic candidates are
reported only by the Belle experiment, and yet to be confirmed by
an independent experiment.

 \begin{figure*}[h]
 \centerline{\includegraphics[width=1\textwidth]{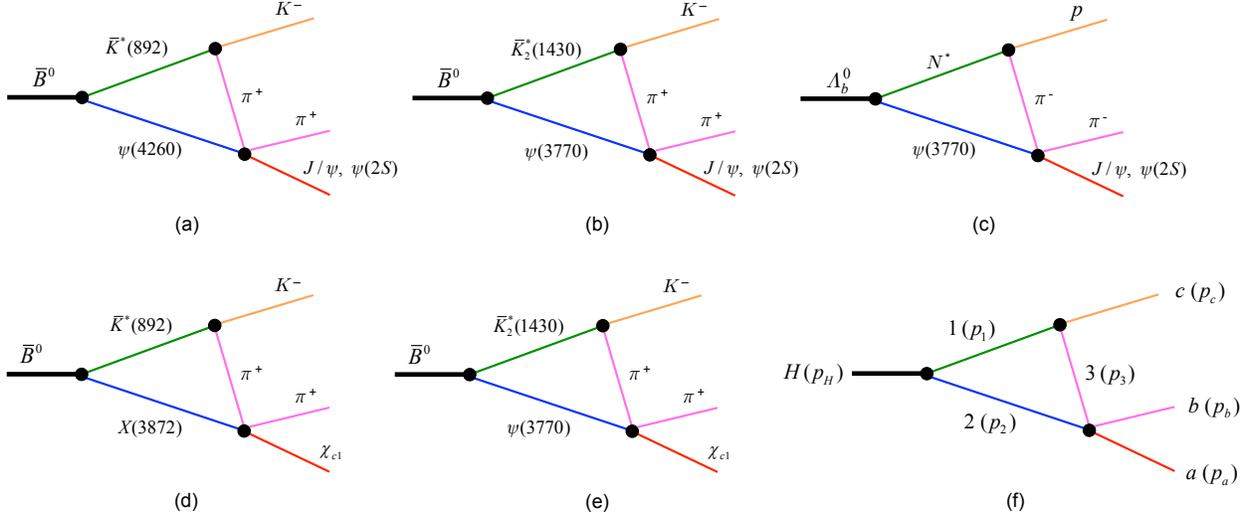}}
  \caption{Triangle diagrams 
  contributing to $\bar B^0\to\psi_f K^-\pi^+$ (a,b),
  $\Lambda_b^0\to\psi_f p\pi^-$ (c),
  and $\bar B^0\to\chi_{c1} K^-\pi^+$ (d,e);
$\psi_f= J/\psi, \psi(2S)$.
  Particle labels and their momenta used in Eq.~(\ref{eq:amp})
  are defined in (f).
  An isospin 1/2 nucleon resonances of 1400$-$1800 MeV is denoted by 
  $N^*$ in (c).
  A $Z_c(4430)$ [$Z_c(4200)$]-like bump in the $\psi_f\pi$ invariant
  mass distribution is generated by TS from the diagram (a) [(b,c)], 
  while a $Z_1(4050)$ [$Z_2(4250)$]-like bump in the $\chi_{c1}\pi^+$ distribution
  by the TS from the diagram (d) [(e)].
  Figures taken from Refs.~\cite{ts1,ts2}.
  Copyright (2019) APS.
 }
\label{fig:diag}
 \end{figure*}
We propose a completely different scenario based on the triangle
singularity (TS)
to interpret these exotic candidates~\cite{ts1,ts2}.
The TS arise in triangle diagrams as shown in Fig.~\ref{fig:diag}
when the processes are kinematically allowed to occur even at the classical level.
The TS is a kinematical effect and its existence and location are
completely determined by the particle masses involved in the process
in the zero-width limit of unstable particles.
Although the unstable particles have finite widths in reality
and thus the TS are somewhat relaxed, the TS still significantly enhance
the amplitudes. 
In what follows, we demonstrate that the TS arising from 
Fig.~\ref{fig:diag}
induce a resonance-like behavior,
with which we can consistently understand 
the experimentally determined properties of 
 $Z_c(4430)$, $Z_c(4200)$, $Z_1(4050)$ and $Z_2(4250)$
 such as spin-parity, mass, width, and Argand plot.

 \section{Model}
\label{sec:model}

 In our model,
 the amplitude for a triangle diagram in Fig.~\ref{fig:diag}
 is given by 
\begin{eqnarray}
 T_{abc,H} &=& \int d\bm{p}_1\,
  { v_{ab;23}(\bm{p}_a,\bm{p}_b;\bm{p}_2,\bm{p}_3)\,
  \over
  E - E_2(\bm{p}_2) - E_3(\bm{p}_3) - E_c(\bm{p}_c)
  }
  \Gamma_{3c,1}(\bm{p}_3,\bm{p}_c;\bm{p}_1)
%  \nonumber \\
 %
%  &\times&
  { 1  \over
  E - E_1(\bm{p}_1) - E_2(\bm{p}_2) }
  \Gamma_{12,H}(\bm{p}_1,\bm{p}_2;\bm{p}_H)
  \ ,
  \label{eq:amp}
\end{eqnarray}
where we have used the particle labels and their momenta defined in
Fig.~\ref{fig:diag}(f).
The total energy in the center-of-mass (CM) frame is denoted by $E$,
while the energy of a particle $x$ is
$E_x(\bm{p}_x)=\sqrt{\bm{p}^2_x+m^2_x}$
with the mass $m_x$ and momentum $\bm{p}_x$.
For unstable intermediate particles 1 and 2, we use
$E_j(\bm{p}_j)=\sqrt{\bm{p}^2_j + m_j^2} - i\Gamma_j/2\ (j=1,2)$
%$E_j(\bm{p}_j)=m_j + \bm{p}^2_j/2m_j - i\Gamma_j/2\ (j=1,2)$
where $\Gamma_j$ is the width.
We use the mass and width values from Ref.~\cite{pdg}.
In Eq.~(\ref{eq:amp}),
the decay of an unstable particle $R$ to lighter particle-pair $i$-$j$
is described by a vertex $\Gamma_{ij,R}$
and the $23\to ab$ rescattering by $v_{ab;23}$.
We use an $s$-wave interaction of $v_{ab;23}$ for 
Fig.~\ref{fig:diag}(a-d) to be consistent with the experimentally
determined spin-parity of $Z_c(4430)$ and $Z_c(4200)$: $J^P=1^+$.
The spin-parity of $Z_1(4050)$ has not been experimentally determined,
and our model predict it to be $J^P=1^-$.
For Fig.~\ref{fig:diag}(e)
where the intrinsic parity is different between the 23 and $ab$ pairs,
we use two types of  $v_{ab;23}$ 
 from the $s$-wave pair to the $p$-wave pair ($J^P=1^+$ for $Z_2(4050)$),
and vice versa ($J^P=1^-$).

 \section{Results for $Z_c(4430)$ and $Z_c(4200)$}

 \begin{figure*}[h]
\centerline{\includegraphics[width=1\textwidth]{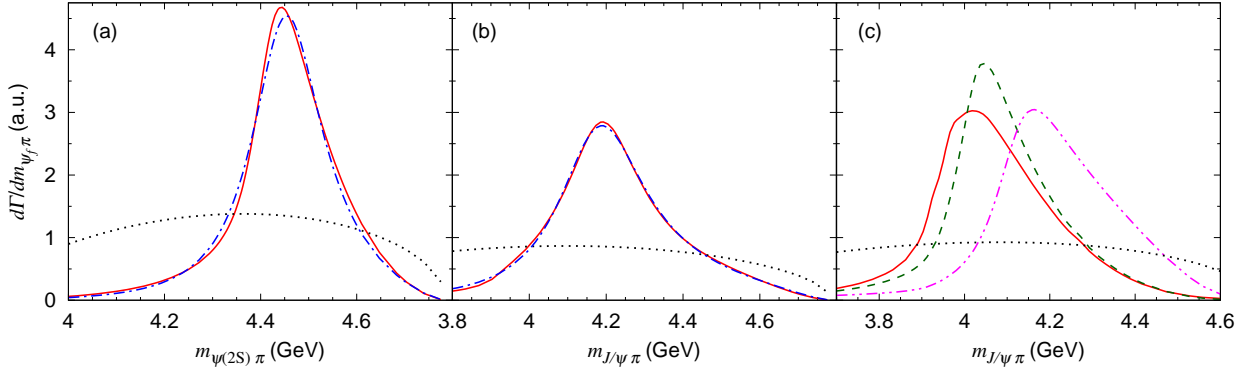}}
\caption{
  The $\psi_f\pi$ ($\psi_f= J/\psi, \psi(2S)$) invariant mass
  distributions of
$\bar B^0\to\psi(2S) K^-\pi^+$ (a),
$\bar B^0\to J/\psi K^-\pi^+$ (b),
and $\Lambda_b^0\to J/\psi p\pi^-$~(c).
  Triangle diagrams Figs.~\ref{fig:diag}(a) and \ref{fig:diag}(b)
  generate the red solid curves in panels (a) and (b), respectively.
  Breit-Wigner amplitudes fitted to the red solid curves are shown by
 the blue dash-dotted curves.
Regarding panel (c),
the triangle diagram of Fig.~\ref{fig:diag}(c) with $N^*=N(1440)\,1/2^+$, $N(1520)\,3/2^-$, and 
  $N(1680)\,5/2^+$ generates
the red solid, green dashed, and magenta dash-two-dotted curves, respectively.
  The phase-space distributions are given by the dotted curves.
  The normalization of each curve, except for the blue dash-dotted ones,
 is fixed to give unity when integrated with respect to
$m_{\psi_f\pi}$.
  Figure taken from Ref.~\cite{ts1}.
    Copyright (2019) APS.
}
\label{fig:spec}
 \end{figure*}
The $\psi(2S)\pi$ [$J/\psi\pi$] invariant mass distribution
for $\bar B^0\to\psi(2S) K^-\pi^+$
[$\bar B^0\to J/\psi K^-\pi^+$]
is shown in Fig.~\ref{fig:spec}(a) [\ref{fig:spec}(b)].
The triangle diagrams of Figs.~\ref{fig:diag}(a) and \ref{fig:diag}(b)
generate the red solid curves in Figs.~\ref{fig:spec}(a) and
\ref{fig:spec}(b), respectively.
The phase-space distributions are also shown by the black dotted curves
for comparison.
Because of the presence of the TS in the triangle diagram, 
a resonance-like peak clearly shows up at
$m_{\psi(2S)\pi}\sim 4.45$~GeV in Fig.~\ref{fig:spec}(a)
($m_{J/\psi\pi}\sim 4.2$~GeV in Fig.~\ref{fig:spec}(b)).
We simulate the peaks from the TS in terms of 
$Z_c^+$ excitations.
We use a model that goes as $\bar B^0\to Z_c^+ K^-$
followed by $Z_c^+ \to \psi_f \pi^+$ to fit 
the Dalitz plot distributions generated by the triangle diagrams
of Figs.~\ref{fig:diag}(a) and \ref{fig:diag}(b).
The Breit-Wigner form is used to model the $Z_c$ propagation.
The kinematical region included in the fit have 
the Dalitz plot distribution larger than 10\% of the peak height. 
As shown by the blue dash-dotted curves in Figs.~\ref{fig:spec}(a) and
\ref{fig:spec}(b), 
the Breit-Wigner form can fit the peaks very well.
\begin{table}[b]
 \caption{\label{tab:BW_param}
Breit-Wigner mass ($M_{BW}$) and width ($\Gamma_{BW}$)
 for $Z_c(4430)$ and $Z_c(4200)$.
 $Z_c(4430)$ [$Z_c(4200)$]
 parameters are obtained by fitting the
 $\bar B^0\to\psi(2S) K^-\pi^+$ (a) [$\bar B^0\to J/\psi K^-\pi^+$ (b)]
  Dalitz plot distributions from
the triangle diagram of Fig.~\ref{fig:diag}(a) [\ref{fig:diag}(b)].
The parameter ranges are from changing the cutoff (vertex form factors) over 0.5--2~GeV.
 The experimentally determined parameters are shown with
the first  statistical (second systematic) errors.
 }
 \setlength{\tabcolsep}{12pt}
\renewcommand\arraystretch{1.3}
\begin{tabular}{cccc|cc}\hline
&\multicolumn{3}{c|}{$Z_c(4430)$}&\multicolumn{2}{c}{$Z_c(4200)$} \\
 & (a) &Belle~\cite{belle_z4430}&LHCb~\cite{lhcb_z4430}
      & (b) &Belle~\cite{belle_z4200} \\\hline
$M_{BW}$ (MeV) & $4463\pm 13$ &  $4485\pm 22^{+28}_{-11}$ & $4475\pm 7^{+15}_{-25}$& $4233\pm 48$  &  $4196^{+31}_{-29}{}^{+17}_{-13}$\\
$\Gamma_{BW}$ (MeV) & $195\pm 16$  & $200^{+41}_{-46}{}^{+26}_{-35}$ & $172\pm 13^{+37}_{-34}$& $292\pm 56$  & $370\pm 70^{+70}_{-132}$\\\hline
%
% $M_{BW}$ (MeV) & 4457 &  $4485\pm 22^{+28}_{-11}$ & $4475\pm 7^{+15}_{-25}$& 4185  &  $4196^{+31}_{-29}{}^{+17}_{-13}$\\
%$\Gamma_{BW}$ (MeV) & 182  & $200^{+41}_{-46}{}^{+26}_{-35}$ & $172\pm 13^{+37}_{-34}$& 276  & $370\pm 70^{+70}_{-132}$\\\hline
\end{tabular}
\end{table}
The resulting Breit-Wigner parameters
are shown in Table~\ref{tab:BW_param}
along with the experimentally determined ones.
The comparison shows a remarkable agreement.

\begin{figure}[h]
\centerline{\includegraphics[width=.45\textwidth]{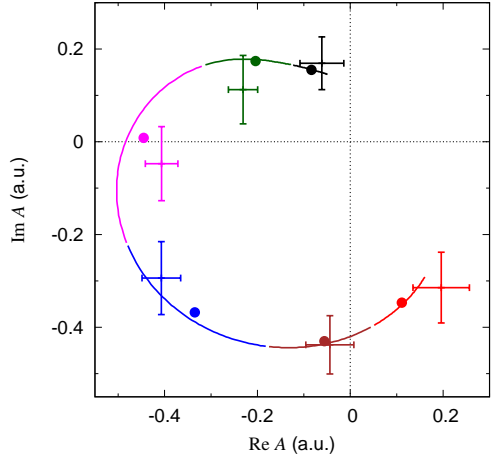}}
\caption{
 Argand plot of the $Z_c(4430)$ amplitude.
The triangle diagram of Fig.~\ref{fig:diag}(a) generates
 six curved segments.
Six data points of the LHCb~\cite{lhcb_z4430}
 are obtained from fitting the data in six bins
 that equally separate
 $18~{\rm GeV}^2\le m^2_{\psi(2S)\pi}\le 21.5~{\rm GeV}^2$; 
 $m^2_{\psi(2S)\pi}$ increases counterclockwise in the figure.
 A curved segment and a data point in the same bin are indicated by
the same color.
 An average [see Eq.~(\ref{eq:average})]
 of a curved segment is shown by the solid circle with 
the same color.
Figure taken from Ref.~\cite{ts1}.
  Copyright (2019) APS.
 }
\label{fig:argand}
\end{figure}
Now we show the triangle amplitude for $Z_c(4430)$
in the form of the Argand plot
so that we can compare it with 
the counterpart from the LHCb~\cite{lhcb_z4430}.
The angle-independent part of the amplitude ($A$)
supplemented with a complex constant background ($c_{\rm \,bg}$)
is given by
\begin{eqnarray}
A(m^2_{ab}) = c_{\rm \,bg} + c_{\rm\,norm} \int d\Omega_{p_c} Y^*_{1,-s^z_{Z_c}}(-\hat{p}_c) M_{abc,H} \ ,
  \label{eq:argad}
\end{eqnarray}
where the $ab$ invariant mass is denoted by $m_{ab}$, and 
the $z$-component of the $Z_c$ spin by $s^z_{Z_c}$;
$Y_{lm}$ is the spherical harmonics.
The quantity $M_{abc,H}$ is
the invariant amplitude related to 
$T_{abc,H}$ of Eq.~(\ref{eq:amp}) through Eq.~(B3) of Ref.~\cite{3pi}.
We adjust complex constants $c_{\rm\,norm}$ and $c_{\rm \,bg}$
to fit the LHCb's Argand plot.
Each point of the LHCb's Argand plot has been fitted to
dataset in the bin covering from $m^2_{\psi(2S)\pi}-\Delta/2$
to $m^2_{\psi(2S)\pi}+\Delta/2$.
Thus we also average our amplitude as:
\begin{eqnarray}
 \bar{A}(m^2_{ab}(i)) = {1\over\Delta}
\int^{m^2_{ab}(i)+\Delta/2}_{m^2_{ab}(i)-\Delta/2} A(m^2_{ab})\, dm^2_{ab} \ ,
  \label{eq:average}
\end{eqnarray}
where the central value of an $i$-th bin is $m^2_{ab}(i)$.
As seen in Fig.~\ref{fig:argand},
the LHCb's $Z_c(4430)$ Argand plot is consistently described by
the triangle diagram of Fig.~\ref{fig:diag}(a) giving
 $\bar{A}(m^2_{ab}(i))$. 
 Thus the counterclockwise behavior of
 the LHCb's $Z_c(4430)$ Argand plot~\cite{lhcb_z4430} is not necessarily
 pointing to the existence of a resonance.

We now discuss $\Lambda_b^0\to J/\psi p\pi^-$.
According to the LHCb analysis~\cite{lhcb_z4200_Lb},
a significantly better description of
the $\Lambda_b^0\to J/\psi p\pi^-$ data is obtained 
by including the $Z_c(4200)$ amplitude.
To this process, 
the triangle diagram of Fig.~\ref{fig:diag}(c) can contribute
with a TS in the $Z_c(4200)$-region.
The triangle diagram includes 
an isospin 1/2 nucleon resonance ($N^*$),
and several $N^*$
in the mass range of 1400$-$1800 MeV can be relevant to the TS.
We show in Fig.~\ref{fig:spec}(c)
the $J/\psi \pi^-$ spectrum generated by
the triangle diagram of Fig.~\ref{fig:diag}(c)
including 
$N^*=N(1440)\,1/2^+$, $N(1520)\,3/2^-$, and $N(1680)\,5/2^+$.
The triangle diagrams with different $N^*$ create
different bumps in the region of $Z_c(4200)$.
In a realistic situation,
a single broad bump from the coherent sum of 
these bumps may show up.
Other charmoniums ($\psi(2S)$, $X(3872)$, etc.)
in the mass range of 3650-3900~MeV, 
which have a coupling to $J/\psi\pi\pi$,
might replace $\psi(3770)$ in Fig.~\ref{fig:spec}(c),
and also generate TS bumps in the $Z_c(4200)$-region.
Because the $\Lambda_b^0\to J/\psi p\pi^-$ data is 
statistically limited,
the $Z_c(4200)$ amplitude in the LHCb analysis
is assumed to have the same mass and width as
the Belle analysis of $\bar B^0\to J/\psi K^-\pi^+$~\cite{belle_z4200}.
Thus, although 
some of the spectrum bumps of Fig.~\ref{fig:spec}(c)
seem to be in the lower end of the $Z_c(4200)$-region,
they are still consistent with the LHCb's observation.

The LHCb analysis~\cite{lhcb_z4200_Lb} also found 
that their description of the $\Lambda_b^0\to J/\psi p\pi^-$ data is
hardly improved by including a $Z_c(4430)$ contribution.
This interesting observation can be understood
if $Z_c(4430)$ appears in $\bar B^0\to\psi(2S) K^-\pi^+$
due to the TS.
This is because, within experimentally observed hadrons, 
there is no triangle diagram like Fig.~\ref{fig:diag}(c) available 
to cause a TS at the $Z_c(4430)$ position for 
the case of $\Lambda_b^0\to J/\psi p\pi^-$.

 \section{Comment on Pakhlov et al.'s triangle diagram}

Pakhlov et al. claimed that a triangle diagram, which includes
an experimentally unobserved hadron, 
can generate a $Z_c(4430)$-like spectrum
bump~\cite{Pakhlov2011,Pakhlov2015} 
due to a kinematical effect.
We however point out that 
the proposed mechanism is
kinematically forbidden at the classical level.
The Coleman-Norton theorem~\cite{coleman} dictates that such a diagram
does not include a TS.
Appropriately substituting the masses, widths, and vertex forms 
into our model discussed in the previous section, we find that
Pakhlov et al.'s triangle diagram does not generate 
a $Z_c(4430)$-like bump, as expected from 
the Coleman-Norton theorem.
The authors presented a clockwise Argand plot from the triangle diagram~\cite{Pakhlov2015}. 
This result has been ruled out by 
the counter-clockwise Argand plot from 
the LHCb~\cite{lhcb_z4430}.
All these points strongly indicate that 
Pakhlov et al.'s scenario is very unlikely to explain
$Z_c(4430)$.

 \section{Results for $Z_1(4050)$ and $Z_2(4250)$}

 \begin{figure*}[h]
\centerline{\includegraphics[width=1\textwidth]{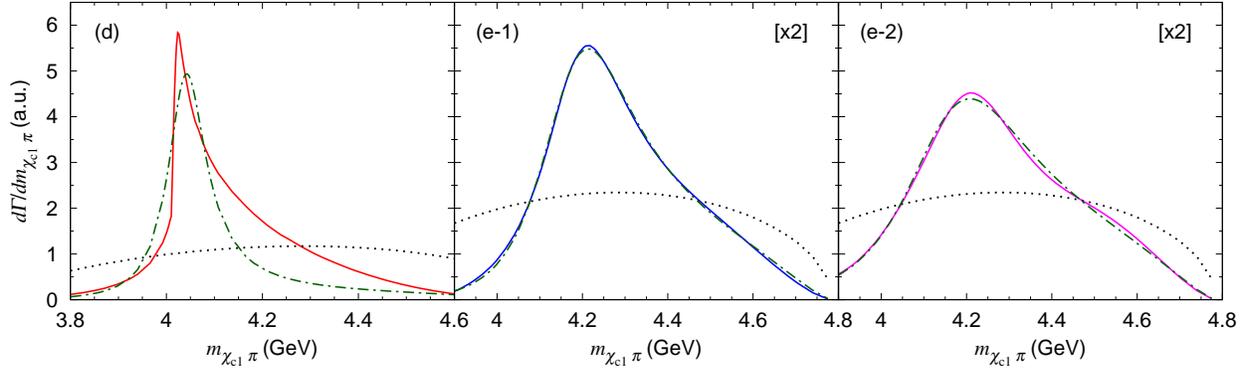}}
\caption{
  Distributions of the $\chi_{c1}\pi^+$ invariant mass for
  $\bar B^0\to\chi_{c1} K^-\pi^+$.
  The red solid curve in the panel (d)
is generated by the triangle diagram of Fig.~\ref{fig:diag}(d).
  The blue [magenta] solid curve in the panel (e-1)
  [(e-2)] is from the diagram Fig.~\ref{fig:diag}(e)
  with the final $\chi_{c1}\pi^+$ pair of $J^P=1^+$ [$J^P=1^-$].
  Breit-Wigner amplitudes, fitted to the solid curves,
  are given with the green dash-dotted curves.
The phase-space distributions are the dotted curves.
  The normalization of the curves has been set in the same way as
Fig.~\ref{fig:spec}.
  The scale has been doubled for the panels (e-1) and (e-2).
Figure taken from Ref.~\cite{ts2}. Copyright (2019) APS.
}
\label{fig:spec2}
 \end{figure*}
The $\chi_{c1}\pi^+$ invariant mass distributions
for $\bar B^0\to\chi_{c1} K^-\pi^+$ are shown in 
Fig.~\ref{fig:spec2}.
The triangle diagram of Fig.~\ref{fig:diag}(d)
gives the red solid curve in Fig.~\ref{fig:spec2}(d).
The diagram Fig.~\ref{fig:diag}(e) generates
the blue solid curve in Fig.~\ref{fig:spec2}(e-1)
for the final $\chi_{c1}\pi^+$ pair with $J^P=1^+$,
while the magenta solid curve in Fig.~\ref{fig:spec2}(e-2)
is the case with $J^P=1^-$.
Clear resonance-like peaks are induced by
the triangle singularities at
$m_{\chi_{c1}\pi}\sim 4.025$~GeV in Fig.~\ref{fig:spec2}(d) and
$m_{\chi_{c1}\pi}\sim 4.22$~GeV in Figs.~\ref{fig:spec2}(e-1) and
\ref{fig:spec2}(e-2).
A characteristic feature of the bump in Fig.~\ref{fig:spec2}(d)
is that it has a significantly asymmetric shape.

\begin{table}[b]
 \caption{\label{tab:BW_param2}
 Breit-Wigner mass ($M_{BW}$), width ($\Gamma_{BW}$),
 and spin-parity ($J^P$) for $Z_1(4050)$ and $Z_2(4250)$.
The Breit-Wigner parameters for 
 $Z_1(4050)$ [$Z_2(4250)$] are determined to fit
 the Dalitz plot distributions for 
$\bar B^0\to\chi_{c1} K^-\pi^+$
 from the triangle diagram of Fig.~\ref{fig:diag}(d) [\ref{fig:diag}(e)].
The parameter ranges are from changing the cutoff (vertex form factors) over 1--2~GeV.
 The Belle analysis result~\cite{belle_z4050}
 on the parameters are shown with 
the first statistical (second systematic) errors.
}
 \setlength{\tabcolsep}{12pt}
\renewcommand\arraystretch{1.3}
\begin{tabular}{ccc|ccc}\hline
&\multicolumn{2}{c|}{$Z_1(4050)$}&\multicolumn{3}{c}{$Z_2(4250)$} \\
 & Fig.~\ref{fig:diag}(d) &Belle~\cite{belle_z4050}
      & \multicolumn{2}{c}{Fig.~\ref{fig:diag}(e)} &Belle~\cite{belle_z4050} \\\hline
$J^P$ & $1^-$ &  $?^?$ & $1^+$ & $1^-$  & $?^?$\\
$M_{BW}$ (MeV) & $4041 \pm 1$ &  $4051\pm 14^{+20}_{-41}$ & $4247 \pm 53$ &$4309 \pm 116$&  $4248^{+44}_{-29}{}^{+180}_{-35}$\\
$\Gamma_{BW}$ (MeV) &  $115 \pm 17$  & $82^{+21}_{-17}{}^{+47}_{-22}$ & $345 \pm 67$&$468 \pm 90$ & $177^{+54}_{-39}{}^{+316}_{-61}$\\\hline
\end{tabular}
\end{table}

We can again simulate the TS-induced bumps with the fake $Z_c$-excitation mechanisms.
The $\bar B^0\to\chi_{c1} K^-\pi^+$
Dalitz plot distribution from the triangle diagram of 
Fig.~\ref{fig:diag}(d) [\ref{fig:diag}(e)] 
is fitted with the mechanism of
$\bar B^0\to Z K^-$ followed by $Z \to\chi_{c1}\pi^+$
by adjusting their Breit-Wigner mass and width.
We include the kinematical region where 
the Dalitz plot distribution is larger than 10\% of the peak height. 
The green dash-dotted curves in Fig.~\ref{fig:spec} are showing the
quality of the fits. 
The Breit-Wigner form cannot fit well
the red solid curve with the asymmetric bump
in Fig.~\ref{fig:spec2}(d).
Meanwhile, 
the bumps in Figs.~\ref{fig:spec2}(e-1) and \ref{fig:spec2}(e-2) are
reasonably well fitted.
The resulting Breit-Wigner parameters are given 
in Table~\ref{tab:BW_param}, 
along with the Belle analysis~\cite{belle_z4050} on
$Z_1(4050)$ and $Z_2(4250)$.
A quite good agreement is seen for $Z_1(4050)$.
Also, our result from the triangle diagram of 
Fig.~\ref{fig:diag}(e) easily agrees with 
the $Z_2(4250)$ mass and width from the Belle analysis
because they have rather large errors.
The $J^P=1^-$ assignment to $Z_2(4250)$ cannot be eliminated by this
comparison alone.

\begin{figure}[h]
\centerline{\includegraphics[width=.48\textwidth]{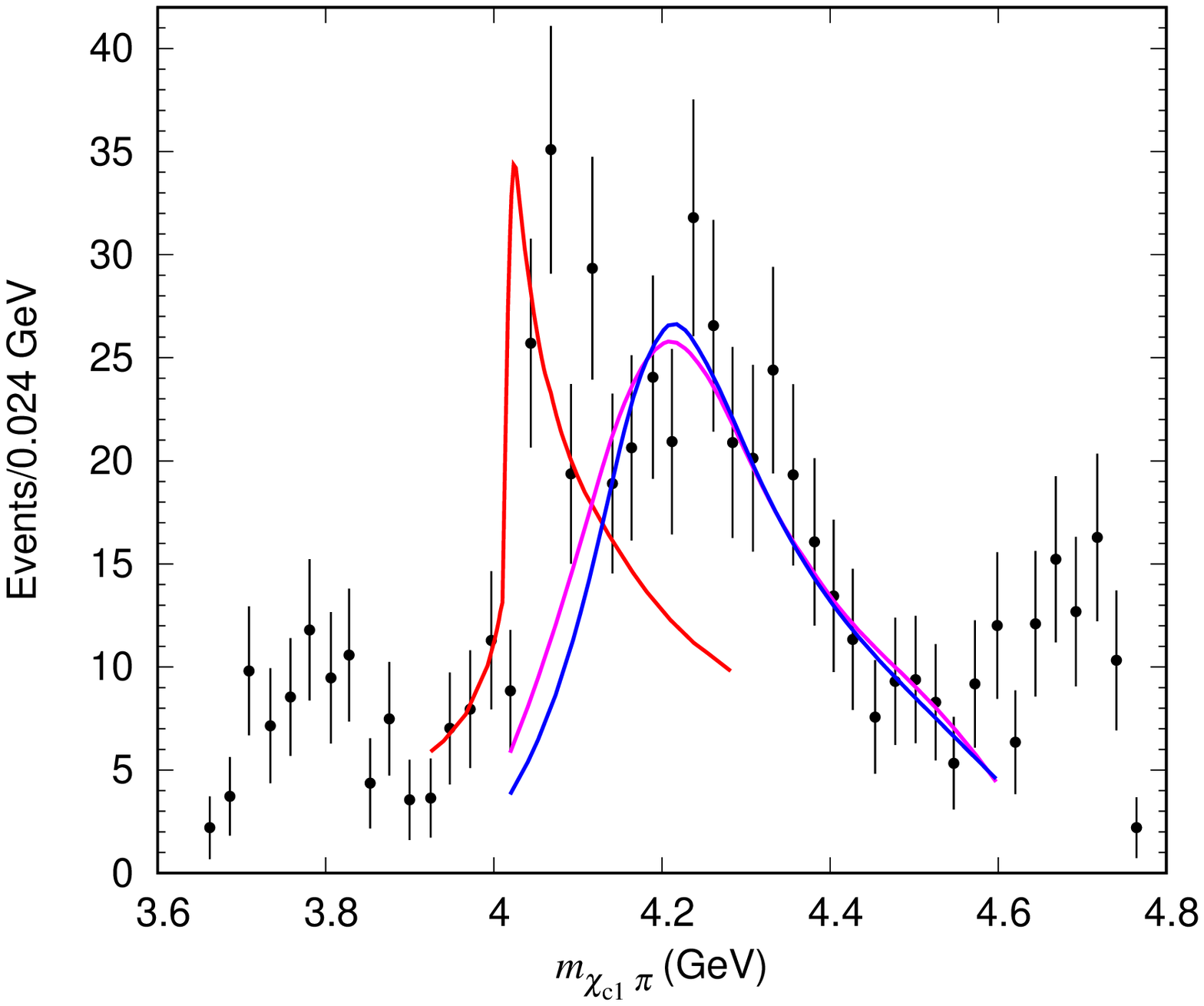}
\includegraphics[width=.49\textwidth]{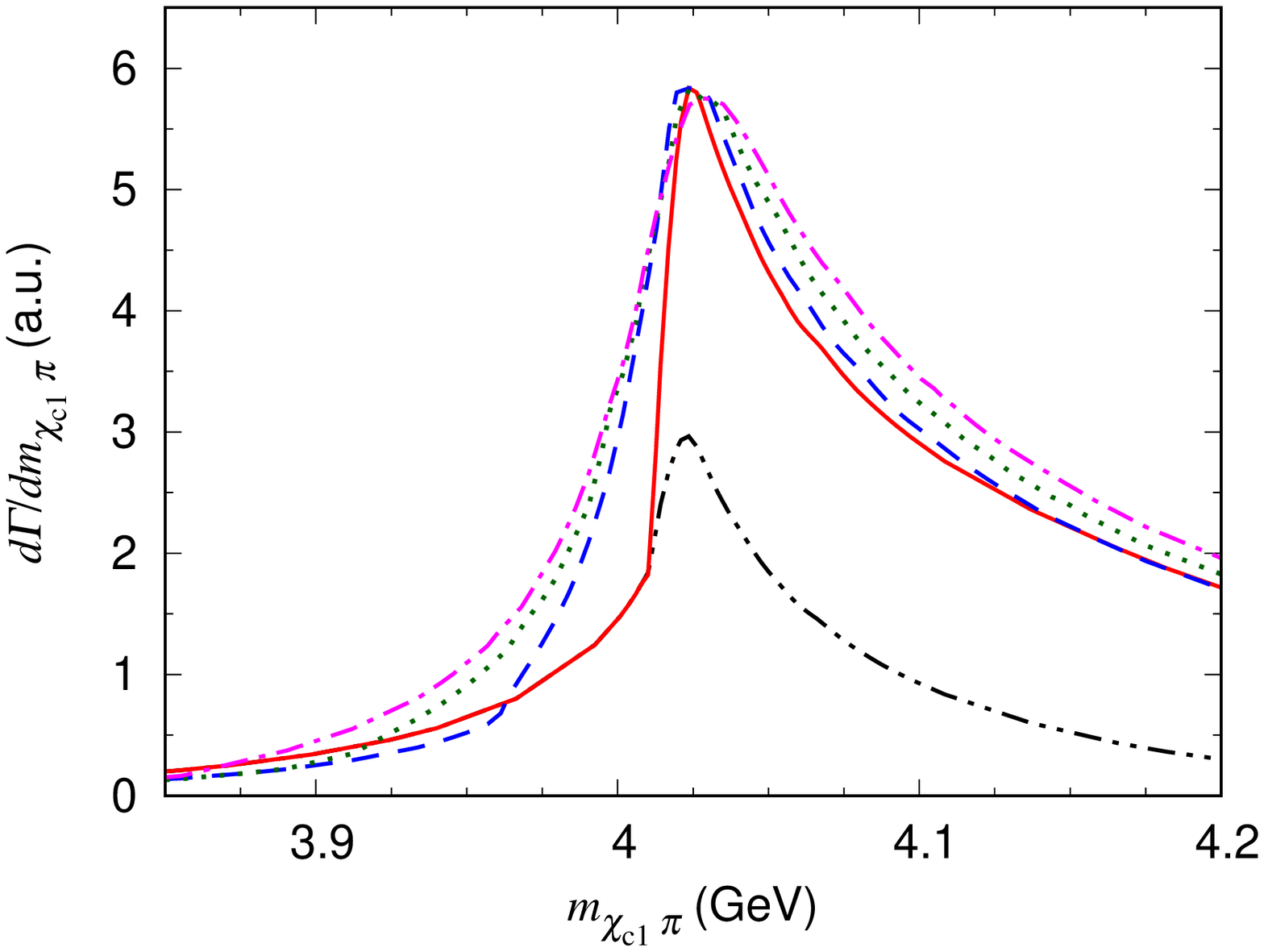}}
\caption{
 Distributions of the $\chi_{c1}\pi^+$ invariant mass
 for $\bar B^0\to\chi_{c1} K^-\pi^+$.
(Left) The red, blue, and magenta solid curves in 
 Figs.~\ref{fig:spec2}(d,e-1,e-2) are superimposed on the Belle data
(Fig.~14 of Ref.~\cite{belle_z4050});
 in order to fit the data,
 a constant factor is multiplied to each of the curves
 and an incoherent constant background is added.
 (Right) The different spectra are generated by 
 the triangle diagram of Fig.~\ref{fig:diag}(e) with different masses for
 $X(3872)$ and $K^*(892)$ (see the text).
We obtain the red solid, blue dashed, green dotted, and magenta dash-dotted curves
using the $X(3872)\pi^+$ threshold energy smaller than the
 PDG value by 0, 50, 100, and 150~MeV, respectively.
 All these curves have the same peak height after being scaled.
 We turn off the on-shell $X(3872)\pi^+$ contribution in
 the red solid curve to obtain
the black dash-two-dotted curve.
 Figures taken from Ref.~\cite{ts2}.
  Copyright (2019) APS.
 }
\label{fig:spec_data}
\end{figure}
The asymmetry of the bump shape generated by 
the triangle diagram Fig.~\ref{fig:diag}(d)
seems important to reproduce the Belle data
in the $Z_1(4050)$-region~\cite{belle_z4050}.
To make this statement clear,
as in Fig.~\ref{fig:spec_data}(left),
we superimpose the spectra from the triangle diagrams of
Figs.~\ref{fig:diag}(d) and ~\ref{fig:diag}(e)
on the Belle data (Fig.~14 of Ref.~\cite{belle_z4050}).
Although this is a qualitative comparison where any interferences among
different mechanisms are not taken into account, 
the spectrum bumps from the triangle diagrams fit the data very well.
Particularly, the data has a very sharp rise and a moderate
fall-off at the $Z_1(4050)$-region, and 
the asymmetric bump shape from 
the triangle diagram of Fig.~\ref{fig:diag}(d) reproduces it well. 
The Belle analysis~\cite{belle_z4050} was not able to fit well
this sharp peak of the data as seen in Fig.~14 of the reference,
probably because they used the Breit-Wigner form
to simulate this bump.
The data seem to disfavor the Breit-Wigner shape.
As seen in Fig.~\ref{fig:spec2}(d),
the triangle diagram of Fig.~\ref{fig:diag}(d)
generates the spectrum bump, the shape of which is
significantly different from the Breit-Wigner.

Having seen that the asymmetric shape is crucial to explain
the Belle data, one may wonder how the triangle diagram can create
this peculiar shape.
In Fig.~\ref{fig:spec2}(d), we can find that the spectrum has an abrupt bend at
$m_{\chi_{c1}\pi}\sim 4.01$~GeV, where the $X(3872)\pi^+$ channel opens,
and the sharp rise of the spectrum starts from this point. 
This is more clearly seen in an enlarged one shown
by the red solid curve in Fig.~\ref{fig:spec_data}(right).
This seems to indicate that the sharp rise is partially due to
the opening of the $X(3872)\pi^+$ channel.
To confirm this speculation,
we turn off the on-shell $X(3872)\pi^+$ contribution,
which arises from $+i\epsilon$ in the denominator
of Eq.~(\ref{eq:amp}),
and show the resulting spectrum by 
 the black dash-two-dotted curve
 in Fig.~\ref{fig:spec_data}(right).
 Indeed, the asymmetry of the bump shape is essentially from
 the on-shell $X(3872)\pi^+$ contribution.

The large asymmetry seems to be also due to 
the proximity of the $X(3872)\pi^+$ threshold to the TS energy 
($\sim 4.025$~GeV).
We can examine this point by lowering the $X(3872)\pi^+$ threshold.
We use $X(3872)$ and $K^*(892)$ masses of, in unit of MeV,
$(m_{X(3872)}, m_{K^*(892)})$=(3822, 1084),
(3772, 1218), and (3722, 1330).
In this way, we can lower the 
threshold by 50, 100, and 150 MeV, respectively, 
while the peak position of the spectrum is kept almost at the same place. 
Figure~\ref{fig:spec_data}(right) indicates that, 
as the $X(3872)\pi^+$ threshold is lowered,
the rise of the bump
becomes significantly more moderate.
Through the above analysis, we now understand
the asymmetric shape of the $Z_1(4050)$ bump observed in
the Belle data
with well-founded physics: TS and
the channel opening near the TS energy.
The triangle diagram of Fig.~\ref{fig:diag}(d) includes 
these physical contents.

The asymmetric bump shape associated with $Z_1(4050)$ is interesting
because it could
sensitively discriminate different theoretical interpretations of $Z_1(4050)$.
A successful model should explain
this characteristic spectrum shape of $Z_1(4050)$,
in addition to the mass, width, and $J^P$.
So far, this question has been successfully addressed by 
our model only.
Higher statistics data is also highly hoped to establish the spectrum shape 
because the error bars are still rather large in the Belle data.

\section{Summary}

The identity of the charged charmonium-like state ($Z_c$) candidates
is a hot problem in the field of the hadron spectroscopy, 
and this work is along this trend.
We showed that the experimentally determined properties of 
 $Z_c(4430)$, $Z_c(4200)$, $Z_1(4050)$ and $Z_2(4250)$
 such as spin-parity, mass, width, and Argand plot
 are all explained well by
  the triangle loop diagrams we identified
and the kinematical singularities involved.
 This scenario is completely different from the previous (and surviving)
 theoretical interpretations based on the four-quark picture
 (including hadron molecule), and is so far the only one giving a
 natural explanation for:
(i) the appearance (absence) of $Z_c(4200)$($Z_c(4430)$)-like
contribution in  $\Lambda_b^0\to J/\psi p\pi^-$;
(ii) the highly asymmetric shape of the $Z_1(4050)$ bump.

% Sections that will go in second font

% Acknowledgement
\section{ACKNOWLEDGMENTS}
The author thanks K. Tsushima for collaboration.
This work is in part supported by 
National Natural Science Foundation of China (NSFC) under contracts 11625523.

% References

\nocite{*}
\bibliographystyle{aipnum-cp}%
%\bibliography{sample}%

\end{document}